\documentstyle[12pt]{article}
\newcommand{\be}{\begin{equation}}
\newcommand{\ee}{\end{equation}}
\newcommand{\bea}{\begin{eqnarray}}
\newcommand{\eea}{\end{eqnarray}}
\newcommand{\al}{\alpha}

\newcommand{\gm}{\gamma}
\newcommand{\Gm}{\Gamma}

\newcommand{\eps}{\epsilon}
\newcommand{\ep}{\epsilon}

\newcommand{\ka}{\kappa}

\newcommand{\om}{\omega}

\newcommand{\dd}{\mbox{d}}

\newcommand{\uQ}{\underline{Q}}
\newcommand{\uM}{\underline{M}}

\newcommand{\uq}{\underline{q}}

\newcommand{\um}{\underline{m}}

\newcommand{\nn}{\nonumber}

\begin{document}
\parindent=1.5pc
\setlength {\unitlength}{1mm}
\begin{titlepage}
\rightline{hep-th/9608151}
\rightline{August 1996}
\bigskip
\begin{center}
{{\bf
Asymptotic Expansions of Feynman Diagrams on the Mass Shell
in Momenta and Masses }\\
\vglue 5pt
\vglue 1.0cm
{ {\large V.A. Smirnov}\footnote{E-mail: smirnov@theory.npi.msu.su } }\\
\baselineskip=14pt
\vspace{2mm}
{\it Nuclear Physics Institute of Moscow State University}\\
{\it Moscow 119899, Russia}
\vglue 0.8cm
{Abstract}}
\end{center}
\vglue 0.3cm
{\rightskip=3pc
 \leftskip=3pc
\noindent
Explicit formulae for asymptotic expansions of Feynman diagrams
in typical limits of momenta and masses with
external legs on the mass shell are presented.
\vglue 0.8cm}
\end{titlepage}

\section{Introduction}

Explicit formulae for asymptotic expansion of Feynman diagrams
in various limits of momenta and masses have been obtained
in the simplest form (with coefficients homogeneous in large
momenta and masses)
in \cite{Go,Ch1,Sm1} (see also \cite{Sm2} for a brief review).
They hold at least when external momenta are off the mass shell.
In the large mass limit, one can also apply the same `off-shell'
formulae for any values of the external momenta.
If however some of the large external momenta are on the mass shell
these formulae are generally non-valid.
It is the purpose of this paper to present
explicit formulae for asymptotic expansions of Feynman diagrams
in two typical limits of momenta and masses with
external legs on the mass shell:
the limit of the large momenta on the mass shell with the large mass and
the Sudakov limit, with the large momenta on the massless mass shell.

To derive these formulae we shall follow a method applied in
ref.~\cite{AZ} for operator product expansions within momentum
subtractions and later for diagrammatic and operator expansions
within dimensional regularization and renormalization \cite{Sm1}.
This method starts with
constructing a remainder of the expansion and using then diagrammatic
Zimmermann identities.

In the next section we shall remind explicit formulae of asymptotic
expansions for the off-shell limit  of large momenta and masses and
illustrate how the remainder of the expansion is constructed.
In Section 3 we shall generalize these formulae for the on-shell
limit of large momenta and masses and, in Section 5, for the Sudakov limit.
These formulae will be illustrated through one-loop examples.

\section{Off-shell limit of large momenta and masses}

It was an idea of Zimmermann \cite{Zi1} to derive operator
product expansion using subtractions of leading asymptotics.
Anikin and Zavialov have systematically developed this idea
\cite{AZ} and constructed a remainder of the operator product
expansions within BPHZ scheme in such a way that it has the
same combinatorial structure as the $R$-operation
(i.e. renormalization at the diagrammatical level) itself.
Let us consider asymptotic expansion of a Feynman diagram
$F_{\Gm} (\uQ , \uq, \uM , \um)$ corresponding to a graph $F_{\Gm}$
in the limit when
the momenta $\uQ \equiv \{ Q_1, \ldots Q_i, \dots \}$ and
the masses $\uM \equiv \{ M_1, \ldots M_i, \dots \}$ are larger than
$\uq \equiv \{ q_1, \ldots q_i, \dots \}$ and
$\um \equiv \{ m_1, \ldots m_i, \dots \}$. Suppose for simplicity
that the external momenta are non-exceptional.
To obtain explicitly the asymptotic expansion
with homogeneity of coefficients in the large momenta and masses, one
uses \cite{Sm1} the same strategy and constructs the remainder in the
form ${\cal R} F_{\Gm} (\uQ , \uq, \uM , \um)$,
with an operation ${\cal R}$ which can be represented, e.g., by the
forest formula \cite{Zi2}
\be
{\cal R} = \sum_F \prod_{\gm \in F} {\cal M}_{\gm} \, .
\label{ff}
\ee
Here the sum runs over forests (sets of non-overlapping subgraphs)
composed of one-particle-irreducible (1PI) or
asymptotically irreducible (AI) subgraphs and
${\cal M}_{\gm} $ is a (pre-)subtraction operator.
Note that for the usual $R$-operation that removes ultraviolet
divergences similar sum is over UV-divergent 1PI subgraphs.
Subtractions in 1PI subgraphs remove UV divergences while
(pre-)subtractions in AI subgraphs remove first terms of the asymptotic
behaviour and
provide desirable asymptotic behaviour of the remainder
in the considered limit.
Thus, if an initial diagram has divergences then the corresponding
operation ${\cal R}$ includes usual UV counterterms. One can consider
asymptotic expansion of regularized diagrams: in this case ${\cal R}$
involves only pre-subtractions. Let us suppose, for simplicity,
that the initial diagram is UV and IR finite.

The class of AI subgraphs is determined by the consider limit.
For example, in the off-shell limit of momenta and masses a subgraph
$\gm$ is called AI if

(a) in $\gm$ there is a path between any pair of external vertices
associated with the large external momenta $Q_i$;

(b) $\gm$ contains all the lines with the large masses;

(c) every connectivity component $\gm_j$ of
the graph $\hat{\gm}$ obtained from $\gamma$ by collapsing
all the external vertices with the large external momenta to a point
is 1PI with respect to the lines with the small masses.

Note that, generally, $\gm$ can be disconnected. One can distinguish
the connectivity component $\gm_0$ that contains external vertices with
the large momenta.

The pre-subtraction operator is nothing but the Taylor expansion
operator in the small momenta and masses:
\be
{\cal M}_{\gm} = {\cal T}^{a_{\gm}}_{\uq^{\gm}, \um^{\gm}} \, .
\label{pso}
\ee
The operator ${\cal T}$ performs Taylor
expansion in the corresponding
set of variables; $\uq^{\gm}$ are the small external momenta of the
subgraph $\gamma$ (i.e. all its external momenta apart from
the large external momenta of $\Gamma$); $\um^{\gamma}$ is a set of
small masses of $\gamma$. The degrees of subtraction $a_{\gm}$
are chosen as $a_{\gm} = \om_{\gm} +\overline{a}$ where
$\om_{\gm}$ is the UV degree of divergence and $\overline{a}$
is the number of oversubtractions. The asymptotic behaviour of the
remainder is governed by the number $\overline{a}$.
Note that these operators are by definition applied
to integrands of Feynman integrals in loop momenta.

To derive the asymptotic expansion it is sufficient then
to write down the identity
\[
F_{\Gm}  =  (1-{\cal R}) F_{\Gm} + {\cal R} F_{\Gm}
\]
and then the diagrammatic Zimmermann identity \cite{Zi2} for the difference
$(1-{\cal R})$, i.e the difference of two $R$-operations.
(Remember that if the initial diagram was renormalized
we would obtain the difference $R-{\cal R}$.)
In our case, this identity takes the form
\be
1 - {\cal R}_{\Gm}
= \sum_{\gm} {\cal R}_{\Gm/\gm} {\cal M}^{a_{\gm}}_{\gm} \, .
\label{ZI}
\ee
where $\Gm/\gm$ are reduced graphs and the sum is over AI subgraphs.

Taking then the number of oversubtractions
$\overline{a}\to \infty$ and using the fact that product of
two operators corresponding to two different AI subgraphs
gives zero (so that ${\cal R}_{\Gm/\gm}$ in (\ref{ZI}) can be replaced
by unity)
one arrives at an explicit formula for the asymptotic expansion:
\be
F_{\Gm} (\uQ,\uq,\uM,\um;\ep)
\; \stackrel{\mbox{\footnotesize$\uM \to \infty$}}{\mbox{\Large$\sim$}} \;
\sum_{\gamma} F_{\Gamma / \gamma} (\uq,\um;\ep)
\circ {\cal T}_{\uq^{\gm}, \um^{\gm}}
F_{\gamma} (\uQ, \uq^{\gm}, \uM,\um^{\gm};\ep) \, ,
\label{LMME}
\ee
where the sum is again in AI subgraphs and the symbol $\circ$
denotes the insertion of the polynomial that stands to the right of it
into the reduced vertex of the reduced diagram $\Gm / \gm$.

Note that individual terms in (\ref{LMME}) possess UV and IR
divergences which are mutually cancelled (see \cite{Ch1,Sm1,Sm2}
for details).

\section{Large momentum expansion on the mass shell}

Let us again consider the large momentum and mass expansion of
a Feynman integral $F_{\Gm}$ corresponding to a graph
$\Gm$. We suppose that the external momenta are non-exceptional, as above.
Moreover, we imply that $\left(\sum_{i\in I} Q_i\right)^2 \neq M_j^2$,
for any subset of indices $I$.
Let now the large external momenta be on the mass shell, $Q_j^2=M_j^2$.
To obtain explicit formulae of the corresponding asymptotic expansion
let us again start from the remainder ${\cal R} F_{\Gm}$ where $\cal R$
is given by (\ref{ff}), with {\em the same} class of AI subgraphs
but a different subtraction operator $\cal M$.

As for the off-shell limit the operator ${\cal M}_{\gm}$ happens
to be a product $\prod_i {\cal M}_{\gm_i}$ of
operators of Taylor expansion in certain momenta and masses.
For connectivity components $\gm_i$ other than $\gm_0$
(this is the component with the large external momenta), the
corresponding operator performs Taylor expansion of the Feynman integral
$F_{\gm_i}$ in its small masses and external momenta. (Note that its small
external momenta are generally not only the small external momenta of the
original Feynman integral but as well the loop momenta of $\Gm$.)
Consider now ${\cal M}_{\gm_0}$. The component $\gm_0$ can be naturally
represented as a union of its 1PI components and cut lines
(after a cut line is removed the subgraph becomes disconnected;
here they are of course lines with the large masses). By definition
${\cal M}_{\gm_0}$ is again factorized and the Taylor expansion of
the 1PI components of $\gm_0$ is performed as in the case of c-components
$\gm_i, \; i=1,2,\ldots$.

It suffices now to describe the action of the
operator ${\cal M}$ on the cut lines. Let $l$ be such a line,
with a large mass $M_j$, and let its momentum be $P_l+k_l$ where
$P_l$ is a linear combination of the large external momenta and
$k_l$ is a linear combination of the loop momenta and small external momenta.
If $P_l=Q_i$ then the operator $\cal M$ for this component of $\gm$ is
\be
\left. {\cal T}_{\ka} \frac{1}{\ka k_l^2+2Q_i k_l +i0} \right|_{\ka=1} \, .
\label{T2}
\ee
(In what follows we shall omit $i0$, for brevity.)
In all other cases, e.g. when $P_l =0$, or it is a sum of two or more
external momenta, the Taylor operator ${\cal M}$ reduces
to ordinary Taylor expansion in small (with  respect to this line
considered as a subgraph) external momenta, i.e.
\be
\left. {\cal T}_{k} \frac{1}{ ( k_l+P_l)^2-M_i^2} \equiv
{\cal T}_{\ka} \frac{1}{ (\ka k_l+P_l)^2-M_i^2} \right|_{\ka=1} \, .
\ee

Note that in all cases apart from the cut lines with $P_l^2 =M^2_j$
the action of the corresponding operator $\cal M$ is graphically
described (as for the off-shell limit) by contraction of the
corresponding subgraph to a point and
insertion of the resulting polynomial into the reduced vertex of the
reduced graph.

Using the same Zimmermann identity (\ref{ZI})
(note however that the symbol $\Gm/\gm$ in ${\cal R}_{\Gm/\gm}$ does
not now have its proper meaning) with the new pre-subtraction
operator we now obtain the following explicit form
of the asymptotic expansion in the on-shell limit
\be
F_{\Gm} (\uQ, \uM, \uq, \um;\ep)
\; \stackrel{\mbox{\footnotesize$M_j \to \infty$}}{\mbox{\Large$\sim$}} \;
\sum_{\gamma}  {\cal M}_{\gm}
F_{\Gm} (\uQ, \uM, \uq, \um;\ep) \, .
\label{eae}
\ee

Let us apply this general formula to a
one-loop propagator-type diagram consisting of two lines,
with masses $M$ and $m$, and an external momentum $p$ with $p^2=M^2$:
\be
\int \dd^d k \frac{1}{(k^2-2pk)(k^2-m^2)} .
\label{FI1}
\ee

In the limit $M\to\infty$, general formula (\ref{eae})
gives contributions corresponding to two subgraphs:
the subgraph $\gm$ associated with the heavy mass $M$ and the
graph $\Gm$ itself. The contribution of $\Gm$ is obtained as
formal Taylor expansion of the propagator with the mass $m$ at
$m=0$:
\be
\sum_{j=0}^{\infty} (m^2)^j
\int \dd^d k \frac{1}{(k^2-2pk)(k^2)^{j+1}} .
\label{F1}
\ee
The integral involved is easily calculated by the $\al$-representation.
We have
\be
i \pi^{d/2} (M^2)^{-\eps}
\sum_{j=0}^{\infty} (-1)^j
\frac{\Gm(j+\eps)\Gm(-2j-2\eps+1)}{\Gm(-j-2\eps+2)} (m^2/M^2)^j .
\label{CGm}
\ee

According to prescription (\ref{T2}) the contribution of
$\gm$ comes from the formal Taylor
expansion of another factor in the integrand $1/(k^2-2pk)$
with respect to $k^2$:
\be
\sum_{j=0}^{\infty} (-1)^j
\int \dd^d k \frac{(k^2)^j}{(-2pk)^{j+1} (k^2-m^2)} .
\label{F2}
\ee
Calculating the above one-loop integral gives
\be
i \pi^{d/2} \frac{1}{2} (m^2)^{-\eps}
\sum_{j=0}^{\infty}
\frac{\Gm((j+1)/2)\Gm((j-1)/2+\eps)}{j!} (m^2/M^2)^{(j+1)/2} .
\label{Cgm}
\ee

In the sum of two contributions artificial IR and UV poles are
cancelled, with the following result:
\bea
i \pi^2 \left\{ \frac{1}{\eps} + 2 - \ln M^2
+ \frac{1}{2} \frac{m^2}{M^2}
\left( \ln \frac{M^2}{m^2} +2 \right) \right. \nn \\
\left. + \frac{1}{2}\frac{m}{M} \sum_{n=0}^{\infty}
\frac{\Gm(n+1/2)\Gm(n-1/2)}{(2n)!} (m^2/M^2)^n
- \frac{1}{2}\frac{m^4}{M^4} \sum_{n=0}^{\infty}
\frac{(n+1)!n!}{(2n+3)!} (m^2/M^2)^n \right\} .
\label{1-l}
\eea
The UV pole which is present from the very
beginning can be removed by usual renormalization.

\section{Explicit formulae for the Sudakov limit}

The large momentum limit of the last section can be considered
either in Euclidean or Minkowski space. An example of typically
Minkowskian situation which has no analogues in Euclidean space
is the Sudakov limit \cite{Sud}. One of its version is formulated
as the behaviour of a three-point Feynman diagram $F_{\Gm}(p_1,p_2,m)$
depending on two momenta, $p_1$ and $p_2$, on the massless mass shell,
$p_i^2=0$, with $q^2\equiv (p_1-p_2)^2 \to -\infty$.
We suppose, for simplicity, that there is one small non-zero mass, $m$.
Let us enumerate three end-points of the diagram according to
the following order: $p_1, p_2, q$.

To treat this limit let us use the previous strategy and construct
a remainder determined by an appropriate pre-subtraction operator and
then apply Zimmermann identities.

Now, we call a subgraph $\gm$ of $\Gm$ AI if at least one of the following
conditions holds:

({\em i}) In $\gm$ there is a path between the end-points 1 and 3.
The graph $\hat{\gm}$ obtained from $\gamma$ by identifying
the vertices 1 and 3 is 1PI.

({\em ii}) Similar condition with $1 \leftrightarrow 2$.

Note that when $q^2$ gets large the components at least of $p_1$ or $p_2$
are large.

The pre-subtraction operator ${\cal M}_{\gm}$ is now naturally
defined as a product $\prod_j {\cal M}_{\gm_j}$ of
operators of Taylor expansion acting on 1PI components and cut lines
of the subgraph $\gm$.
Suppose that the above condition ({\em i}) holds and ({\em ii})
does not hold.
Let $\gm_j$ be a 1PI component of $\gm$ and let $p_1+k$ be one
of its external momenta,
where $k$ is a linear combination of the loop momenta.
(We imply that the loop momenta are chosen
in such a way that $p_1$ flows through all $\gm_j$ and the
corresponding cut lines).
Let now $\uq_j$ be other independent external momenta of $\gm_j$.
Then the operator $\cal M$ for this component is
defined as\footnote{The formulated prescription for the pre-subtraction
operator works at least up to two-loop level. Its justification
and clarification for the general situation will
be given in future publications.}
\be
{\cal T}_{k-((p_1 k)/(p_1p_2)) p_2,\uq_j, \um_j } \, ,
\label{T2a0}
\ee
where $\um_j$ are the masses of $\gm_j$.
In other words, it is the operator of Taylor expansion in $\uq_j$ and $\um_j$
at the origin and in $k$ at the point
$\tilde{k} =\frac{(p_1 k)}{(p_1p_2)} p_2$ (which depends on $k$ itself).

For the cut lines we adopt the same prescription.
With $p_1 +k$ as the momentum of the cut line, we have
\be
\left. {\cal T}_{\ka} \frac{1}{\ka (k_l^2-m_l^2) +2 p_1 k }
\right|_{\ka=1} \, ,
\label{T2a}
\ee
which is similar to (\ref{T2}).
If both conditions ({\em i}) and ({\em ii}) hold the corresponding
operator performs Taylor expansion in the mass and the external momenta
of subgraphs (apart from $p_1$ and $p_2$).
For example, ${\cal M}_{\gm}$ for the whole graph ($\gm=\Gm$)
gives nothing but Taylor expansion of the integrand in $m$.

Consequently we arrive at an explicit expansion similar to (\ref{eae}):
\be
F_{\Gm}(p_1,p_2,m,\ep)
\; \stackrel{\mbox{\footnotesize$q^2 \to -\infty$}}{\mbox{\Large$\sim$}} \;
\sum_{\gamma}  {\cal M}_{\gm}
F_{\Gm}(p_1,p_2,m,\ep) \, ,
\label{eae1}
\ee
with the new operator ${\cal M}_{\gm}$ and the new definition of
AI subgraphs.

\begin{figure}[h]
\setlength {\unitlength}{1mm}
\begin{picture}(130,60)(0,0)
\put (50,10) {\line(3,4){24}}
\put (50,10) {\line(1,0){48}}
\put (98,10) {\line(-3,4){24}}
\put (74,42) {\line(0,1){14}}
\put (50,10) {\line(-3,-2){14}}
\put (98,10) {\line(3,-2){14}}
\put (36,8) {\makebox(0,0)[bl]{$p_2$}}
\put (107,8) {\makebox(0,0)[bl]{$p_1$}}
\put (78,48) {\makebox(0,0)[bl]{$q=p_1-p_2$}}
\put (73,3) {\makebox(0,0)[bl]{$m$}}
\put (91,26) {\makebox(0,0)[bl]{$m_2=0$}}
\put (44,26) {\makebox(0,0)[bl]{$m_1=0$}}
\end{picture}
\caption{ }
\end{figure}
Let us illustrate these prescriptions through an example of a one-loop
triangle Feynman integral, with the masses $m_1=m_2=0, m_3=m$ and
the external momenta $p_1^2=p_2^2=0, q=p_1-p_2$ (see Fig.~1)
\be
F_{\Gm}(p_1,p_2,m,\ep)
= \int \frac{\dd^dk}{(k^2-2 p_1 k) (k^2-2 p_2 k) (k^2-m^2)} .
\label{triangle}
\ee
The set of AI subgraphs consists of two single massless lines as well
as the graph $\Gm$ itself, with the corresponding pre-subtraction
operators ${\cal M}^a_1, {\cal M}^a_2$ and ${\cal M}^a_0$.
The remainder of the corresponding asymptotic expansion is
${\cal R}^{a} F_{\Gm}$ where
\be
{\cal R}^{a} = (1-{\cal M}^a_0) (1 -{\cal M}^a_1 -{\cal M}^a_2 )
\label{remainder}
\ee
and the pre-subtraction operators ${\cal M}^a_i$ are
\bea
{\cal M}^a_0 \frac{1}{k^2-m^2}
= \sum_{j=0}^a \frac{(m^2)^j}{(k^2)^{j+1}} \, ;
\label{Ma0} \\
{\cal M}^a_i \frac{1}{k^2 - 2 p_i k}
= \sum_{j=0}^a \frac{(k^2)^j}{(- 2 p_i k)^{j+1}} \, , \;\; i=1,2 \, .
\label{Ma12}
\eea
It is implied that each of these operators acts only on the
corresponding factor of the integrand and does not touch other
two factors.
The remainder is UV and IR finite, for any $a$. Its asymptotic
behaviour is $(m^2)^{a+1}/(k^2)^{a+2}$ modulo logarithms.

Then the terms of the expansion result from the Zimmermann
identity:
\be
1 = (1 - {\cal R}) + {\cal R} =
{\cal M}_0 + {\cal M}_1 + {\cal M}_2 + \ldots
\ee
where we have dropped zero products of different operators and
turned to the limit $a\to\infty$
(with ${\cal M}_i = {\cal M}^{\infty}_i$).

All the resulting one-loop integrals are easily evaluated for any order
of the expansion.
The operator ${\cal M}_0$ gives the following contribution
at $\eps\neq 0$:
\be
{\cal M}_0 F_{\Gm} = -i \pi^{d/2} \frac{1}{(-q^2)^{1+\eps}}
\sum_{n=0}^{\infty}
\frac{\Gm(n+1+\eps)\Gm(-n-\eps)^2}{\Gm(1-n-2\eps)}
\left(\frac{m^2}{q^2}\right)^n \, .
\label{M00}
\ee
The terms ${\cal M}_1$ and ${\cal M}_2$ are not individually
regulated by dimensional regularization but their sum exists
for general $\eps\neq 0$:
\bea
\left( {\cal M}_1 + {\cal M}_2 \right) F_{\Gm} = i \pi^{d/2}
\frac{1}{q^2 (m^2)^{\eps}} \Gm(\eps) \Gm(1-\eps)
\sum_{n=0}^{\infty} \frac{1}{n! \Gm(1-n-\eps)}
\nn \\ \times
\left[ \ln (-q^2/m^2) + \psi(\eps) + 2\psi(n+1)- \psi(1)- \psi(1-\eps)
- \psi (1-n-\eps) \right]
\left(\frac{m^2}{q^2}\right)^n \, .
\label{M012}
\eea
Here $\psi$ is the logarithmic derivative of the gamma function.

Taking into account terms with arbitrary $j$ in (\ref{Ma0})
and (\ref{Ma12}) and calculating the corresponding one-loop
integrals we come, in the limit $\eps\to 0$, to the known result:
\bea
F_{\Gm} \left. \right|_{\eps=0}
\; \stackrel{\mbox{\footnotesize$q^2 \to -\infty$}}{\mbox{\Large$\sim$}}
\left( {\cal M}_0 + {\cal M}_1 + {\cal M}_2 \right)
F_{\Gm} \left. \right|_{\eps=0} \nn \\
= - \frac{i\pi^2}{q^2} \left[
\mbox{Li}_2\left(\frac{1}{t}\right) + \frac{1}{2} \ln^2 t
- \ln t \ln(t-1) -\frac{1}{3} \pi^2 \right] \, ,
\label{tri-res}
\eea
where Li$_2$ is the dilogarithm and $t=-q^2/m^2$.

\section{Conclusion}

The explicit formulae of the on-shell asymptotic expansions
presented above can be successfully applied for calculation of
Feynman diagrams --- see, e.g., \cite{ACVS} where the formulae of
Section~3 are used.  It looks also natural to apply the formulae
for the Sudakov limit for analyzing asymptotic behaviour of
multiloop diagrams.

Observe that any term of asymptotic expansions (\ref{eae},\ref{eae1}),
in particular (\ref{1-l},\ref{tri-res}), is calculated much easier than
the whole diagram itself\footnote{The present work hence solves the
problem, raised by \cite{Co1} and discussed in \cite{Sm3}, of how to
expand (\ref{triangle}) {\em without} knowing the full result.}.
For example, an {\em arbitrary} term of the expansion of a
2-loop on-shell diagram considered in \cite{ACVS}
can be in principle analytically evaluated by computer.
Let us note that the formulae
for the Sudakov limit give not only the leading asymptotic
behaviour (``the leading twist'') but any power with respect to
the expansion parameter and all the logarithms:  in the
considered one-loop examples all the powers and logarithms were
obtained at the same footing.
Thus it looks reasonable to apply the presented technique to
extend well-known results  on asymptotic behaviour in the Sudakov
limit \cite{Sud,Sud1,Co,Kor} (see also references in \cite{Co})
and obtain all powers and logarithms, at least at the 2-loop
level.

\vspace{2mm}

This research has been supported by the Russian Foundation for Basic
Research, project 96--01--00654.
\vspace{2mm}

{\em Acknowledgments.}
I am very much grateful to K.G.~Chetyrkin, A.~Czarnecki,
K.~Melnikov and J.B.~Tausk for helpful discussions.

\end{document}